\documentclass[aps,prd,twocolumn,groupedaddress,nofootinbib,balancelastpage]{revtex4-1}

\usepackage{amssymb}
\usepackage{graphicx}
\usepackage{amsmath}
\usepackage{hyperref}
\usepackage{xspace}

\hypersetup{linktocpage} 
\hypersetup{
    colorlinks,
    citecolor=blue,
    filecolor=blue,
    linkcolor=blue,
    urlcolor=blue}

\newcommand{\as}{\alpha_s}

\newcommand{\C}{{\cal C}}
\newcommand{\G}{{\cal G}}

\newcommand{\kt}{{\vec{k}_t}}
\newcommand{\qt}{{\vec{q}_t}}
\newcommand{\m}{{\cal M}}

\newcommand{\beq}{\begin{equation}}
\newcommand{\eeq}{\end{equation}}
\newcommand{\bea}{\begin{eqnarray}}
\newcommand{\eea}{\end{eqnarray}}
\newcommand{\bdm}{\begin{displaymath}}
\newcommand{\edm}{\end{displaymath}}
\def\as{\alpha_s}

\def\m{{\cal M}}

\def\ord{{\cal O}}

\def \G{\mathcal{G}}

\def\d{\text{d}}

\def \msb{\overline{\textrm{MS}}}

\begin{document}

\title{Combining $Q_T$ and small-$x$ resummations}

\author{Simone Marzani}
\email{smarzani@buffalo.edu}
\affiliation{University at Buffalo, The State University of New York, Buffalo NY 14260-1500, USA}


\begin{abstract}
We analyze transverse momentum ($Q_T$) resummation of a colorless final state, e.g.\ Higgs production in gluon fusion or the production of a lepton pair via the Drell-Yan mechanism, in the limit where the invariant mass of the final state is much less than the center-of-mass energy, i.e. $Q^2\ll s$. 
We show how the traditional resummation of logarithms of $Q_T/Q$ can be supplemented with the resummation of the leading logarithmic contributions at small $x=Q^2/s$ and we compute the necessary ingredients to perform such joint resummation.
\end{abstract}

\maketitle
\section{Introduction}\label{sec:intro}

Early this year the CERN Large Hadron Collider (LHC) has resumed its operations at 13~TeV.
Thanks to this increase in the colliding energy, searches for new particles can be pushed to higher and higher masses.
The primary goal of the LHC Run II remains the precision study of physics at the electro-weak scale. One key element in this rich physics program is the measurement of the Higgs production cross section and differential distributions~\cite{Aad:2014lwa,Aad:2014tca,Aad:2015lha,CMS:2015hja} and their comparison to state-of-the-art theoretical predictions. The hierarchy between the protons' center-of-mass energy and the electroweak scale $s\gg Q^2 \sim m_h^2$, where $m_h$ is the Higgs mass, induces potentially large logarithmic corrections in their ratio $x=\frac{Q^2}{s}\ll1$, which can, and should, be resummed to all orders in perturbation theory~\cite{Lipatov:1976zz,Fadin:1975cb,Kuraev:1976ge,Kuraev:1977fs,Balitsky:1978ic,Fadin:1998py}.

Moreover, precision measurements of the production cross section for lepton pairs, via the Drell-Yan (DY) process, as a function of the pair invariant mass $Q$, give us a unique opportunity to explore the vast kinematic plane covered by the LHC, and they provide us with a clean probe of parton distribution functions (PDFs). In particular, the LHCb detector is ideally suited to probe the low-$x$ region~\cite{LHCb:2012fja}.

One of the most important measurements both in the context of Higgs and DY is the boson's transverse momentum ($Q_T$) distribution. It is well known that while in the region $Q_T\sim Q$ fixed-order perturbation theory can be trusted, the description of the perturbative region $Q_T\ll Q$ (with $Q_T\gg \Lambda_\text{QCD}$) requires all-order resummation. It follows that if we consider the Higgs or DY low-$Q_T$ region at the LHC, we are in the presence of a double hierarchy of scales, namely $Q_T/Q\ll 1$ and $x\ll 1$. The aim of this current study is to outline a formalism to simultaneously resum logarithms of both ratios. 

Examples of joint resummations already appeared in the literature. For instance, Refs.~\cite{Li:1998is,Laenen:2000ij,Kulesza:2002rh,Kulesza:2003wn} considered the simultaneous resummation of $Q_T/Q$ and threshold, i.e. large-$x$, logarithms. Furthermore, a framework to consistently combine small- and large-$x$ resummation for inclusive cross-sections was proposed in Ref.~\cite{Ball:2013bra}. Examples of simultaneous resummation of two observables have been also discussed in the context of Soft-Collinear Effective Theory, e.g.~\cite{Procura:2014cba,Larkoski:2015kga}.

In this study, we show that the traditional resummation of logarithms of $Q_T/Q$~\cite{Collins:1984kg}, which is currently known to next-to-next-to leading logarithmic accuracy (NNLL), can be easily supplemented with the resummation of small-$x$ logarithmic contributions. As we shall discuss in the following, both coefficient functions and parton density evolution kernels resum at small-$x$. 
The consistent treatment we are going to employ  is to resum coefficient functions to their leading (non-trivial) order (LL$x$), while
resumming DGLAP evolution kernels to NLL$x$. 
The double-resummed result is a rather simple modification of the usual $Q_T$ resummation formula, essentially because the logarithmic enhancements have different kinematic origins. We will denote the accuracy of the joint resummation with NNLL+LL$x$.

A study of small-$x$ contributions in the context of the DY process, for both invariant mass and $Q_T$ distributions, was performed in Refs.~\cite{deOliveira:2012ji,deOliveira:2012mj}, where the use of DY as a probe of the low-$x$ region of the PDFs was advocated. In that study the all-order inclusion of small-$x$ effects was mimicked by a phenomenologically motivated choice of the factorization scale for DGLAP evolution. As stated above, in this current study we instead  aim to achieve a well-defined formal accuracy.

Our findings appear to be in agreement with a previous analysis~\cite{Mueller:2012uf,Mueller:2013wwa}, which considered the resummation of $Q_T$ logarithms (referred in that work as ``Sudakov logarithms") in the context of the small-$x$ saturation formalism. Our derivation follows an orthogonal logic in that we start from $Q_T$-resummed expression and we aim to supplement it with small-$x$ resummation.

In this paper we limit ourselves to derive the general framework for the joint resummation, while leaving a detailed phenomenological analysis to an upcoming dedicated study.
We start our discussion with a short recap of $Q_T$ resummation in Sec.~\ref{sec:smallQT} and of small-$x$ resummation in Sec.~\ref{sec:smallx}, while our main result is derived in Sec.~\ref{sec:double}, before concluding in Sec.~\ref{sec:conclusions}. The calculation of the DY $Q_T$ distribution is summarized in the Appendix.

\section{A recap of $Q_T$ resummation} \label{sec:smallQT}

In this section we summarize the main ingredients of $Q_T$ resummation for an electro-weak final state, i.e.\ Higgs or DY. For simplicity, we are going to consider distributions which are fully inclusive in the electro-weak boson decay products, as well as integrated in the boson's rapidity. Extension to more differential distributions~\cite{deFlorian:2012mx,Grazzini:2013mca,Catani:2015vma} is possible and will be investigated elsewhere. 

The literature on $Q_T$ resummation is vast and since the seminal paper Ref.~\cite{Collins:1984kg}, there has been a continuous effort in producing accurate theoretical predictions that can describe experimental data. For example, high logarithmic accuracy~\cite{Bozzi:2003jy,Bozzi:2005wk,Bozzi:2010xn,Catani:2010pd,Becher:2010tm,Becher:2012yn,Catani:2013tia,Gehrmann:2014yya,Neill:2015roa} has been achieved and computer programs that allow one to compute NNLL predictions matched to next-to-leading order (NLO) for the $Q_T$ distribution in case of colorless final states in hadron collision exist, e.g.~\cite{Ladinsky:1993zn,Landry:2002ix,Bozzi:2005wk,Bozzi:2010xn,Banfi:2012du,deFlorian:2011xf,deFlorian:2012mx}. 
Moreover, NNLO accuracy has recently been achieved~\cite{Boughezal:2015aha,Boughezal:2015dra,Boughezal:2015dva,Boughezal:2013uia,Chen:2014gva,Ridder:2015dxa}, paving the way for N$^3$LL resummation. 

Moreover, observables such as $\phi^*$~\cite{Vesterinen:2008hx,Banfi:2010cf} that exploit angular correlations to probe similar physics as $Q_T$, while being measured with much better experimental resolution, have triggered theoretical studies to extend the formalism of $Q_T$ resummation to these new variables~\cite{Banfi:2009dy,Banfi:2011dm,Banfi:2011dx,Banfi:2012du,Guzzi:2013aja}. The experimental resolution of $\phi^*$  is so good~\cite{Abazov:2010mk,Aaij:2012mda,Aad:2012wfa,Aad:2014xaa,Abazov:2014mza} that the theoretical uncertainty of the state-of-the-art NNLL+NLO calculation is much larger than the experimental one, calling for improved theoretical predictions. In particular, the results of Ref.~\cite{Abazov:2010mk} showed that some small-$x$ models~\cite{Nadolsky:2000ky,Berge:2004nt} performed poorly when compared to the data, demanding a deeper understanding of small-$x$ effects in the low-$Q_T$ region.
 
$Q_T$ resummation is usually performed in Fourier space and $b$ is the variable conjugated to $Q_T$. The small-$Q_T$ region corresponds to large $b$ and logarithms of $Q_T$ are mapped into logarithms of $1/b$.
Following Refs.~\cite{Catani:2000vq,Catani:2010pd} we write the resummed transverse momentum distribution for the production of an electroweak final state $F$ from initial-state partons $c$ and  $\bar{c}$ as
\begin{align}\label{eq:QT_res}
&\frac{\d \sigma}{ \d Q_T^2}=  \sigma^\text{born}_{c\bar{c}\to F} \int \d x_1 \int \d x_1
\int_0^\infty \d b \frac{b}{2} \,J_0(b Q_T) S_c(b, Q)  \nonumber \\ &\times
 \int \d z_1 \int \d z_2\,  \delta \left(1-z_1 z_2 \frac{x_1 x_2 s}{Q^2} \right) 
 \nonumber \\ &\times \Bigg [H_{c \bar{c}}^F\left(\as(Q) \right) 
 C_{c a_1}\left(z_1,\as\left(\tfrac{b_0}{b}\right)\right)  C_{\bar{c} a_2}\left(z_2,\as\left(\tfrac{b_0}{b}\right)\right) 
 \nonumber \\&
 +\tilde{H}_{c \bar{c}}^F\left(\as(Q) \right)
G_{c a_1}\left(z_1,\as\left(\tfrac{b_0}{b}\right)\right)  G_{\bar{c} a_2}\left(z_2,\as\left(\tfrac{b_0}{b}\right) \right)
\Bigg] \nonumber \\ &\times
f_{a_1}\left(x_1,\tfrac{b_0}{b}\right)f_{a_2}\left(x_2,\tfrac{b_0}{b}\right),
\end{align}
where the sum over $a_1,a_2$ is understood.  
For Standard Model Higgs production we have $F=h$, $c=\bar{c}=g$, and $H=\tilde{H}$, while for DY production we have $F=Z/ \gamma^*$ and $c=q$, and $G_{q,a}=G_{\bar{q},a}=0$. We have also introduced the Bessel function $J_0$ and $b_0=2 e^{-\gamma_E}$, where $\gamma_E$ is the Euler constant.

Our aim is to understand, isolate and, eventually resum, the leading high-energy behavior, i.e.\ the small-$x$ regime of the above $Q_T$-resummed expression.
However, Eq.~(\ref{eq:QT_res}) contains several convolution integrals, which make the structure of high-energy singularities rather opaque. We can greatly simplify our analysis by considering another conjugate space. We therefore take Mellin moments of the $Q_T$ distribution with respect to $x$
\begin{equation}
\Sigma(N,Q_T^2)= \int_0^1 \d x \, x^{N-2}\frac{\d \sigma}{ \d Q_T^2},
\end{equation}
obtaining
\begin{align}\label{eq:QT_res_mellin}
&\Sigma(N,Q_T^2)=  \sigma^\text{born}_{c\bar{c}\to F}
\int_0^\infty \d b \frac{b}{2} \,J_0(b Q_T) S_c(b, Q) \nonumber \\ &\times
\Bigg [H_{c \bar{c}}^F\left(\as(Q) \right) \mathcal{C}_{c a_1}\left(N,\as\left(\tfrac{b_0}{b}\right)\right)  \mathcal{C}_{\bar{c} a_2}\left(N,\as\left(\tfrac{b_0}{b}\right)\right) \nonumber \\  &+\tilde{H}_{c \bar{c}}^F\left(\as(Q) \right)
\mathcal{G}_{c  a_1}\left(N,\as\left(\tfrac{b_0}{b}\right)\right)  \mathcal{G}_{\bar{c} a_2}\left(N,\as\left(\tfrac{b_0}{b}\right) \right)
\Bigg] 
\nonumber \\ &\times
\mathcal{F}_{a_1}\left(N,\tfrac{b_0}{b}\right) \mathcal{F}_{a_2}\left(N,\tfrac{b_0}{b}\right),
\end{align}
where calligraphic symbols denote Mellin moments of the original functions, i.e.
\begin{align}
\mathcal{C}_{ab}(N,\as)&= \int_0^1 \d z \, z^{N-1}  C_{ab}(z,\as), \nonumber \\   \mathcal{G}_{ab}(N,\as)&= \int_0^1 \d z \, z^{N-1} G_{ab}(z,\as),\nonumber \\ \mathcal{F}_a(N,\mu)&= \int_0^1\d z \, z^{N-1} f_a(z,\mu). 
\end{align}
Logarithmic corrections in the limit $x\to0$ (or $z\to 0$) correspond in Mellin space to multiple poles at $N=1$.

It is now useful to discuss the different ingredients that enter the $Q_T$ resummation formulae Eq.~(\ref{eq:QT_res}) and~(\ref{eq:QT_res_mellin}), commenting on their universality properties, as well as their behavior in the two limits of interest, i.e. small $Q_T$ (large $b$) and small $x$ ($N\sim1$). The first term that we encounter is the Born cross-section. This is clearly a process-dependent piece, but a trivial one, because the processes we are considering have $2\to 1$ kinematics with no $b$ or $N$ dependence. The Sudakov form factor $S_c$ is the object of primary importance in $Q_T$ resummation. We note that it is only a function of $L=\ln b$ and it resums such contributions to all orders in perturbation theory, to the desired accuracy. Noticeably, as we will show below, this ingredient is not touched by small-$x$ resummation, leading to a particularly simple way of merging the two resummations. 
The hard functions $H$ and $\tilde{H}$ are instead process-dependent and incorporate the contributions from hard virtual corrections (note the scale of the coupling). 
These contributions do not depend on $N$ and they are of the form $1+\ord(\as)$. Therefore, they only contribute beyond LL$x$ i.e.\ outside the scope of this analysis and can be taken at fixed-order. For the processes of interest, they are currently known to two-loop accuracy~\cite{Catani:2011kr,Catani:2012qa}. The functions $C_{ab}$ ($\mathcal{C}_{ab}$) and $G_{ab}$ ($\mathcal{G}_{ab}$) are instead universal and only depend on the parton flavor $a$ and $b$, similar to what happens for the Altarelli-Parisi splitting kernel, to which they are indeed related. By performing explicit calculations in $k_t$-factorization~\cite{Catani:1990eg}, we will determine their LL$x$ behavior.
Finally, the last contribution is given by the PDFs $f_a$ ($\mathcal{F}_a$). As it is well known, the DGLAP kernels, which control the evolution of the PDFs, contain themselves logarithmic contributions at small-$x$, which need to be resummed. This subject has been widely studied in the literature and small-$x$ resummation of the splitting functions can be performed to NLL$x$ accuracy, as we briefly review in the next section.

\section{A recap of small-$x$ resummation} \label{sec:smallx}
In this section we briefly review the resummation of high-energy, or small-$x$, logarithms that appear both in perturbative coefficient functions and in the DGLAP kernels, which govern the evolution of the parton densities.

Small-$x$ resummation of parton evolution is usually performed in Mellin space. It is well known that only one of the two eigenvalues of the singlet anomalous dimension matrix contains, to all orders in $\as$, the contributions with the highest powers of the rightmost $N=1$ singularity and therefore needs to be accounted for to all orders. 
The resummation of these contributions is based on the BFKL equation~\cite{Lipatov:1976zz,Fadin:1975cb,Kuraev:1976ge,Kuraev:1977fs,Balitsky:1978ic,Fadin:1998py}.
However, it turns out that the correct inclusion of LL$x$ and NLL$x$ corrections is far from trivial. This problem received great attention in the 1990s, by more than one group, see, for instance, Refs.~\cite{Salam:1998tj,Ciafaloni:1999yw,Ciafaloni:2003rd,Ciafaloni:2007gf} and Refs.~\cite{Ball:1995vc,Ball:1997vf,Altarelli:2001ji,Altarelli:2003hk,Altarelli:2005ni,Altarelli:2008aj}.

The all-order behavior of the anomalous dimension (which is a simple pole at $N=N_B$, to the right of $N=1$, but close to it) relies on the inclusion of two classes of formally subleading corrections~\cite{Altarelli:2005ni} on top of the NLL$x$ terms: namely, running coupling
corrections, without which the $N$-space leading singularity would be
a square-root cut instead of a simple pole~\cite{Altarelli:2001ji}, and anticollinear terms~\cite{Salam:1998tj} without which the perturbative expansion of both the position and residue of the above simple pole would not be stable. Schematically, the resummed and matched anomalous dimension matrix, in the singlet sector, is~\footnote{Henceforth a plus superscript will denote a small-$x$ resummed and matched expression.}
\begin{align}\label{resum_splitting}
\gamma_{ab}^+(N,\as)&= \gamma_{ab}^\text{NLO}(N,\as)+\gamma_{ab}^\text{NLL$x$}(N,\as)\nonumber \\&- \gamma_{ab}^{\text{NLL}x,\as^2}(N,\as).
\end{align}
where the first term is the fixed-order DGLAP anomalous dimension, the second one includes NLL$x$ contributions~\cite{Fadin:1998py} as well as anticollinear terms and running coupling corrections via the Bateman anomalous dimension~\cite{Altarelli:2005ni} and the last term avoids double counting.

The resummation of partonic coefficient functions is based on the so-called $k_t$-factorization theorem~\cite{Catani:1990xk,Catani:1990eg,Collins:1991ty,Catani:1993ww,Catani:1993rn,Catani:1994sq,Ball:2007ra,Caola:2010kv} and it is known to LL$x$ for an increasing number of cross-sections and distributions~\cite{Ball:2001pq,Hautmann:2002tu,Marzani:2008az,Harlander:2009my,Marzani:2008uh,Diana:2010ef,Caola:2011wq}. Its generalization to rapidity distributions was carried out in Refs~\cite{Caola:2010kv}, where it was applied to the case of Higgs production in gluon fusion.

Very recently, small-$x$ resummation has been also extended to transverse momentum distributions~\cite{Forte:2015gve}. Those results lay the foundations for the present study. 
Analogous to the inclusive case, one computes the leading order transverse momentum distribution for the relevant
process, while keeping the initial-state gluon(s) off their mass-shell. A triple Mellin transform is then computed: an $N$-Mellin, defined as above, and two Mellin transforms with respect to the off-shellness of the incoming gluons $\xi_i=\frac{|k^2|}{Q^2}$, with moments $M_1$ and $M_2$, respectively:
\begin{align} \label{QTimpactfactor}
h(N,M_1,M_2,\xi_p)&= M_1 M_2 \int_0^1 \d z \, z^{N-1} 
\int_0^\infty \d \xi_1 \, \xi_1^{M_1-1} \nonumber \\&\times
\int_0^\infty \d \xi_2 \,  \xi_2^{M_2-1} \frac{\d \sigma^\text{off-shell}}{\d \xi_p}, 
\end{align}
with $\xi_p=\frac{Q_T^2}{Q^2}$. 
The leading (non-trivial) singularity of the partonic coefficient function is obtained by identifying the Mellin variables with the LL$x$ BFKL anomalous dimension $M_i=\gamma_s$, where
\begin{equation}\label{gammas}
\gamma_s =\sum_{n=1}^\infty e_{n} \left(\frac{\as}{N-1} \right)^n,
\end{equation}
where the coefficients $e_{n}$ are determined~\cite{Jaroszewicz:1982gr} using
duality~\cite{Ball:1999sh} from the leading order BFKL kernel. 
Furthermore, it can be shown that the explicit $N$ dependence in Eq.~(\ref{QTimpactfactor}) is subleading.

The result obtained with the above procedure is in a factorization scheme which is often denoted with $Q_0 \msb$. It turns out that this scheme is actually preferable for performing small-$x$ resummation~\cite{Altarelli:2008aj} because of better convergence properties. Furthermore, results in the more commonly used $\msb$ scheme are easily recovered by a multiplicative scheme-change factor $R_{\msb}$~\cite{Catani:1993ww,Catani:1994sq,Ciafaloni:2005cg,Marzani:2007gk}:
\begin{equation}\label{Rmsb}
R_{\msb}(M)=1+\frac{8}{3}\zeta_3 M^3 +\ord \left(M^4\right).
\end{equation}

Finally, we mention the important fact that the inclusion of an all-order class of subleading running coupling corrections in the resummed coefficient function is further required~\cite{Ball:2007ra, Altarelli:2008aj} in order for this not to develop extra spurious singularities. We leave the inclusion of these contributions to a future phenomenological study.

\section{Small-$Q_T$ and small-$x$}  \label{sec:double}
The main result of this paper is an expression which simultaneously resums large logarithms of $Q_T$ and of $x$ to NNLL+LL$x$ accuracy. It has a particularly simple form:
\begin{align}\label{eq:QT_res_smallx_mellin}
&\Sigma^+(N,Q_T^2)=  \sigma^\text{born}_{c\bar{c}\to F}
\int_0^\infty \d b \frac{b}{2} \,J_0(b Q_T) S_c(b, Q) \nonumber\\&\times
\Big [H_{c \bar{c}}^F\left(\as(Q) \right) \mathcal{C}^+_{c  a_1}\left(N,\as\left(\tfrac{b_0}{b}\right)\right)  \mathcal{C}^+_{\bar{c} a_2}\left(N,\as\left(\tfrac{b_0}{b}\right)\right) \nonumber \\  &+\tilde{H}_{c \bar{c}}^F\left(\as(Q) \right)
\mathcal{G}^+_{c  a_1}\left(N,\as\left(\tfrac{b_0}{b}\right)\right)  \mathcal{G}^+_{\bar{c} a_2}\left(N,\as\left(\tfrac{b_0}{b}\right) \right)
\Big] \nonumber\\&\times
 \mathcal{F}^+_{a_1}\left(N,\tfrac{b_0}{b}\right) \mathcal{F}^+_{a_2}\left(N,\tfrac{b_0}{b}\right).
\end{align}
The above NNLL+LL$x$ result has the same form as standard $Q_T$ resummation Eq.~(\ref{eq:QT_res_mellin}), with two new ingredients: resummed (and matched) coefficient functions:
\begin{align}\label{resum_coeff_C}
\C_{ab}^+(N,\as)&= \C_{ab}^\text{NNLO}(N,\as)+\C_{ab}^\text{LL$x$}(N,\as)\nonumber \\&- \C_{ab}^{\text{LL}x,\as^2}(N,\as)  , \\
\G_{ab}^+(N\as)&= \G_{ab}^\text{NNLO}(N,\as)+\G_{ab}^\text{LL$x$}(N,\as)\nonumber \\&- \G_{ab}^{\text{LL}x,\as^2}(N,\as). \label{resum_coeff_G}
\end{align}
and resummed (and matched) PDFs $\mathcal{F}^+_{a}$, i.e. parton densities that evolve with the resummed anomalous dimensions given in Eq.~(\ref{resum_splitting}).
Analogous to Eq.~(\ref{resum_splitting}), the last terms in Eqs.~(\ref{resum_coeff_C}) and~(\ref{resum_coeff_G}) are the expansion of the resummation to NNLO and they avoid double counting. 

In the remainder of this section we are going to derive the above double-resummed expression. We start with (re)deriving the separation between Sudakov and PDF evolution in Sec.~\ref{sec:evolution}, which will allows us to use known results Eq.~(\ref{resum_splitting}) to perform the resummation of DGLAP evolution. We will then move to the resummation of the coefficient functions Eqs.~(\ref{resum_coeff_C}) and~(\ref{resum_coeff_G}) in Sec.~\ref{sec:coefficient_functions}.

\subsection{Sudakov form factor and PDF evolution} \label{sec:evolution}
We now consider the resummation of small-$Q_T$ logarithms, with the aim of isolating potential sources of large logarithmic corrections in $x$, i.e.\ poles in $N=1$. There exist many derivations in the literature, but we find the one in Ref.~\cite{Banfi:2009dy} particularly useful for our purposes. For convenience, we work at NLL, but the analysis goes through to NNLL as well. As a further simplification, we explicitly consider only the flavor-diagonal contributions, while restoring full flavor dependence in the end. 

We compute the cumulative distribution for the transverse momentum of the Higgs (or $Z/\gamma^*$) boson to be less than a given value $Q_T$. 
At Born level $gg \to h$ or $q\bar q \to Z/\gamma^*$, so the boson has no transverse momentum. We then consider the emission of an arbitrary number of collinear gluons off the incoming hard legs. In order to diagonalize the integral, we perform the calculation in the conjugate space $(\vec{b},N)$ by taking Fourier moments with respect to the two-dimensional vector $\vec{Q}_T$ and Mellin moments with respect to $x$:
\begin{align} \label{real_emissions}
W^\text{real}(b,N) &= \sum_{n=0}^\infty \frac{1}{n!}\prod_{i}^n \int [\d k_i] \,z_i^{N-1} (2 C_c) \frac{\as(k_{ti})}{2 \pi}  \nonumber \\& \quad \times \bar{P}^\text{real}(z_i)  e^{i \vec{b} \cdot \vec{k}_{ti}}
\Theta \left(k_{ti}-Q_0 \right) \nonumber \\  &\quad \quad\times  \Theta \left(1-z_i+ \frac{k_{ti}}{Q} \right),
\end{align}
where the emitted gluon phase space is $[\d k_i]=\d z_i \frac{\d k_{ti}^2}{k_{ti}^2}\frac{\d \phi_i}{2 \pi}$ and $C_c=C_F, C_A$ is the appropriate color factor. The first $\Theta$ function expresses the fact that emissions below the cut-off $Q_0$ belong to the non-perturbative region of the proton wave-function, while the second one correctly accounts for the large-angle soft region of phase-space.
In order to achieve NLL accuracy, the strong coupling $\as$ has to be evaluated at two loops, in the CMW scheme~\cite{Catani:1990rr}.

The series in Eq.~(\ref{real_emissions}) sums to an exponential. Analogously, we have also to consider the virtual corrections $W^\text{virtual}$, which do not change the transverse momentum $Q_T$ and trivially exponentiate. 
The emission probability is described by the real and virtual matrix elements (see e.g.\ App.~E of Ref.~\cite{Banfi:2004yd}):
\begin{align}\label{splitting_functions_real}
\bar{P}^\text{real}(z)&= 
\begin{cases}
\frac{1+z^2}{1-z}, \quad \text{for a quark},\\
\frac{2 z}{1-z}+\frac{2(1-z)}{z}+ 2 z(1-z), \, \text{for a gluon};
\end{cases}\\
\bar{P}^\text{virtual}(z)&=(-1) 
\begin{cases}
\frac{1+z^2}{1-z}, \quad \text{for a quark},\\
\frac{2 z}{1-z}+ z(1-z)\\ \,+n_f T_R (z^2+(1-z)^2), \, \text{for a gluon}.
\end{cases} \label{splitting_functions_virtual}
\end{align}
For later convenience, we also introduce the leading order regularized splitting functions
\begin{align}\label{splitting_plus}
P_{qq}(z)&=\frac{\as}{2 \pi} C_F \left[\frac{1+z^2}{1-z}\right]_+, \nonumber \\
P_{gg}(z)&=\frac{\as}{2 \pi} 2 C_A \Bigg[\left(\frac{z}{1-z}+\frac{z(1-z)}{2} \right)_+ \nonumber \\&+\frac{1-z}{z}+\frac{z(1-z)}{2}-\frac{2}{3}n_f T_R \delta(1-z)\Bigg],
\end{align}
and the corresponding anomalous dimensions
\begin{align}
\gamma_{cc}(N,\as) &= \int_0^1 z^{N-1} P_{cc}(z), \quad c=q,g.
\end{align}
Thus, the resummed exponent is obtained by putting together real, virtual and PDF contributions:
\begin{align}\label{real_virtual}
R(b,N)&= -\ln \Big[ W^\text{real}(b,N) W^\text{virtual} W^\text{PDF}(b,N) \Big] \nonumber\\
&=2 C_c \int [\d k] \frac{\as(k_{t})}{2 \pi}  \Theta \left(k_{t}-Q_0 \right)\Theta \left(1-z+ \frac{k_t}{Q} \right)
\nonumber \\ & \times
 \left(- z^{N-1} \bar{P}^\text{real}(z) e^{i \vec{b} \cdot \vec{k}_{t}}-\bar{P}^\text{virtual}(z) \right) 
 \nonumber \\ &
+ 2 \int_{Q_0^2}^{Q^2} \frac{\d k_t^2}{k_t^2} \gamma_{cc}(N,\as(k_t)).
\end{align}
By rewriting $z^{N-1}=1+(z^{N-1}-1)$ and using the definitions in Eqs.~(\ref{splitting_functions_real}), (\ref{splitting_functions_virtual}), and~(\ref{splitting_plus}), we are able to reshuffle the contributions to the resummed exponent as follows
\begin{align}\label{reshuffle}
R(b,N)&
=-\int_{Q_0^2}^{Q^2} \frac{\d k_t^2}{k_t^2} \int_0^{2 \pi} \frac{\d \phi}{2 \pi} \left( 1- e^{i \vec{b} \cdot \vec{k}_{t}} \right)
\nonumber \\ & \times
  \Bigg[ \int_0^{1-\frac{k_t}{Q}} \d z\frac{\as(k_{t})\, C_c}{\pi} \bar{P}^\text{virtual}(z) \nonumber \\&-2 \gamma_{cc}(N,\as(k_t)) \Bigg]  +\ord\left(\frac{k_t}{Q}\right).
\end{align}
The factor $\left( 1- e^{i \vec{b} \cdot \vec{k}_{t}} \right)$ essentially acts as a cut-off on the $k_t$ integral. At NLL we have~\footnote{See Ref.~\cite{Catani:2003zt} for a generalization of this approximation to higher-logarithmic accuracy.}
\begin{align}\label{NLLradiator}
R(b,N)
=&-\int_{b_0^2/b^2}^{Q^2} \frac{\d k_t^2}{k_t^2} \Bigg[ \int_0^{1-\frac{k_t}{Q}} \d z\frac{\as(k_{t})\, C_c}{\pi}\bar{P}^\text{virtual}(z) 
\nonumber\\&
-2 \gamma_{cc}(N,\as(k_t)) \Bigg] \nonumber \\
&= -\ln S_c +2 \int_{b_0^2/b^2}^{Q^2}  \frac{\d k_t^2}{k_t^2} \gamma_{cc}(N,\as(k_t)).
\end{align}
Thus, we have successfully separated two distinct contributions: the Sudakov form factor ($S_c$), computed here at NLL accuracy (and systematically improvable) and a DGLAP contribution, which evolves the PDFs from the hard scale $Q$ down to $b_0/b$. Note that here we have only considered flavor-diagonal splittings. Off-diagonal ones do not alter the Sudakov form factor and they are fully taken into account by the complete DGLAP evolution.

 Noticeably, the Sudakov form factor does not exhibit any $N$ dependence and it is therefore insensitive to small-$x$ enhancements. On the other hand, the DGLAP contribution does contain $N=1$ singularities, which have to be accounted for to all orders. As already discussed, its resummation is known to NLL$x$ and it is given by Eq.~(\ref{resum_splitting}).

\subsection{Coefficient functions} \label{sec:coefficient_functions}
We now turn our attention to the coefficient functions $C_{ab}(z,\as)$ and $G_{ab}(z,\as)$. In fixed-order perturbation theory, they are known to NNLO accuracy for both Higgs~\cite{Catani:2011kr} and vector boson~\cite{Catani:2012qa} production. Let us start by the coefficient function $G$:
\begin{align} \label{G_coefficient_expansion}
G_{qa}(z,\as)&=0, \\
G_{ga}(z,\as)&= \frac{\as}{\pi} G_{ga}^{(1)}(z)+  \ord \left(\as^2 \right),
\end{align}
with
\begin{equation} \label{G_coeff_NLO}
G_{gg}^{(1)}(z) =C_A \frac{1-z}{z}, \quad G_{gq}^{(1)}(z) =C_F \frac{1-z}{z}.
\end{equation}
The separation of the hard function $H$ and the coefficient function $C$ requires the definition of a particular resummation scheme. We follow the convention of Ref.~\cite{Catani:2013tia}. We have
\begin{align} \label{C_coefficient_expansion}
C_{ab}(z,\as)&= \delta_{ab} \delta(1-z) + \frac{\as}{\pi} C_{ab}^{(1)}(z)
\nonumber\\&+\left(\frac{\as}{\pi} \right)^2 C_{ab}^{(2)}(z) +\ord \left(\as^3\right),
\end{align}
with the following NLO coefficients:
\begin{align}
C_{gg}^{(1)}(z)&=0, \quad   C_{gq}^{(1)}(z)= \frac{C_F}{2} z, \label{C_coeff_NLO}\\
C_{qq}^{(1)}(z)&=\frac{C_F}{2} (1-z), \quad   C_{qg}^{(1)}(z)= \frac{1}{2} z (1-z) \label{C_coeff_NLO_DY}.
\end{align}
Note that because QCD conserves flavor $C_{q\bar q}^{(1)}=C_{qq'}^{(1)}=C_{q\bar q'}^{(1)}=0$. The NNLO contributions have fairly lengthy expressions. Here we are interested in their small-$z$ behavior:
\begin{align}
C_{gg}^{(2)}(z)&= 0 +\dots \; , \quad   C_{gq}^{(2)}(z)= 0 +\dots, \label{C_coeff_NNLO}\\
C_{qq}^{(2)}(z)&=\frac{C_F}{z} \left( \frac{43}{108}-\frac{\pi^2}{36}\right)+ \dots \;, \nonumber\\  C_{qg}^{(2)}(z)&=  \frac{C_A}{z} \left( \frac{43}{108}-\frac{\pi^2}{36}\right)+\dots
\label{C_coeff_NNLO_DY}
\end{align}
where the dots denote contributions beyond LL$x$. 
Moreover, the coefficient functions $C_{q\bar q}^{(2)}$, $C_{qq'}^{(2)}$ and $C_{q\bar q'}^{(2)}$ have the same LL$x$ behavior as $C_{qq}^{(2)}$.

In order to determine the all-order LL$x$ behavior of the above coefficient function we make use of the $k_t$-factorization theorem~\cite{Catani:1990xk,Catani:1990eg,Collins:1991ty,Catani:1993rn,Catani:1993ww,Catani:1994sq}, in particular its recent extension to transverse momentum distributions~\cite{Forte:2015gve}. Because of their universal nature, it is sufficient to perform two distinct calculations to extract $C_{ga}$, $G_{ga}$ and $C_{qa}$, respectively, with $a=q,g$. Moreover, we anticipate that the LL$x$ coefficient functions with $a=q$ and $a=g$ will be related by color charge relations~\cite{Catani:1994sq}, as supported by their fixed-order expansions Eq.~(\ref{G_coeff_NLO}), and Eq.~(\ref{C_coeff_NNLO_DY}).

\subsubsection{Gluon coefficient functions}\label{sec:coeff_gluon}
In order to compute $C_{ga}$ and $G_{ga}$ we consider Higgs production in gluon fusion in the framework of $k_t$-factorization.
At this point, one might wonder whether we should consider the calculation in the effective theory (EFT) where the top has been integrated out or the full theory. It is known that the EFT, despite being an excellent approximation to the inclusive cross-section in the full theory (at least at up to  NNLO~\cite{Pak:2009dg,Harlander:2009my}), fails in certain kinematic limits, in particularly at large Higgs transverse momentum~\cite{Baur:1989cm} or at small $x$~\cite{Hautmann:2002tu,Marzani:2008az}. 
However, here we are interested in the LL$x$ behavior at small $Q_T$, hence the full theory and the EFT are equivalent, up to an overall normalization given by the hard function $H$.

Working in the framework of $k_t$-factorization, we then consider the transverse momentum distribution of $gg \to h$, computed at LO in the EFT, with off-shell gluons, which are responsible for a non-vanishing Higgs transverse momentum even at LO. 

According to Eq.~(\ref{QTimpactfactor}), we then consider Mellin moments with respect to $z=m_h^2/\hat{s}$, $\xi_1=|k_1^2|/m_h^2$, and $\xi_2=|k_2^2|/m_h^2$.  The explicit expression for the off-shell cross-section can be found in Ref.~\cite{Forte:2015gve}. Here we are interested in its Fourier transform.   We also normalize out the LO result and we find~\footnote{As already mentioned the explicit $N$ dependence in Eq.~(\ref{QTimpactfactor}) is subleading. Thus, after accounting for the unit shift in the Mellin-space definition of partonic coefficient functions versus partonic cross section, we can safely set $N=0$ in $h(N,M,\xi_p)$.}
\begin{align} \label{Higgs_coeff_prod}
&c(M_1,M_2,b)= \int_0^\infty \d Q_T \, 2 Q_T \,J_0(b Q_T) \, 
 \frac{h\left(0,M_1,M_2,\xi_p\right)}{\sigma^\text{born}} \nonumber \\
&= \left(\frac{b_0}{b m_h}\right)^{2(M_1+M_2)} e^{2\gamma_E(M_1+M_2)}\frac{\Gamma(1+M_1)\Gamma(1+M_2)}{\Gamma(2-M_1)\Gamma(2-M_2)} 
\nonumber \\ &  \quad \quad \quad \quad \quad  \quad \quad \times
\Big[(1-M_1)(1-M_2)+M_1 M_2 \Big].
\end{align}
The $b$-dependent prefactor accounts for the evolution of the PDFs from the hard scale $Q$ down to $b_0/b \sim Q_T$ and therefore corresponds to the contribution discussed in Sec~\ref{sec:evolution}. We note that Eq.~(\ref{Higgs_coeff_prod}) has the structure expected from standard $Q_T$-resummation Eq.~(\ref{eq:QT_res_mellin}), i.e.\ it is  the sum of two contributions, each of which is factorized with respect to the initial-state legs, therefore justifying the double-resummed result Eq.~(\ref{eq:QT_res_smallx_mellin}). We can then read off the Mellin moments of the coefficient function
\begin{align} 
\mathcal{C}_{gg}(M)&= e^{2 \gamma_E M} \frac{\Gamma(1+M)}{\Gamma(1-M)},   \\
\mathcal{G}_{gg}(M)&= e^{2 \gamma_E M} M \frac{\Gamma(1+M)}{\Gamma(2-M)}. 
\end{align}
Moreover, using small-$x$ color-charge relations, we have
\begin{align} \label{gluon_results_gq}
\mathcal{C}_{gq}(M)&= \frac{C_F}{C_A} \left[\mathcal{C}_{gg}(M)-\mathcal{C}_{gg}(0) \right]=  \frac{C_F}{C_A} \left[\mathcal{C}_{gg}(M)-1\right], \\
\mathcal{G}_{gq}(M)&=   \frac{C_F}{C_A} \left[\mathcal{G}_{gg}(M)-\mathcal{G}_{gg}(0) \right]=   \frac{C_F}{C_A} \mathcal{G}_{gg}(M).
\end{align}
The leading high-energy behavior in Mellin space is then found by identifying $M=\gamma_s$ as in Eq.~(\ref{gammas}).

A good check of our calculation can be performed by expanding the resummed results and by comparing them with their fixed-order counterparts. We obtain
\begin{align} \label{gluon_results_gg}
\mathcal{C}_{gg}^{\text{LL}x}(N,\as) &=R_{\msb}(\gamma_s)\, \mathcal{C}_{gg}(\gamma_s)= 1+\ord\left(\frac{\as^3}{(N-1)^3} \right),\\
\mathcal{G}_{gg}^{\text{LL}x}(N,\as)&=R_{\msb}(\gamma_s)\, \mathcal{G}_{gg}(\gamma_s)
\nonumber \\&
= \frac{\as}{\pi} \frac{C_A}{N-1}+ \left(\frac{\as}{\pi} \frac{C_A}{N-1}\right)^2
\nonumber \\&+\ord\left(\frac{\as^3}{(N-1)^3} \right),
\end{align}
and
\begin{align} \label{gluon_results_gg}
\mathcal{C}_{gq}^{\text{LL}x}(N,\as) &=R_{\msb}(\gamma_s)\, \mathcal{C}_{gq}(\gamma_s)= \ord\left(\frac{\as^3}{(N-1)^3} \right),\\
\mathcal{G}_{gq}^{\text{LL}x}(N,\as)&= R_{\msb}(\gamma_s)\, \mathcal{G}_{gq}(\gamma_s)
\nonumber \\&
=\frac{\as}{\pi} \frac{C_F}{N-1}+ \left(\frac{\as}{\pi} \right)^2 \frac{C_F C_A}{(N-1)^2}
\nonumber \\& +\ord\left(\frac{\as^3}{(N-1)^3} \right),
\end{align}
which are in agreement with Eqs.~(\ref{G_coeff_NLO}), (\ref{C_coeff_NLO}), and~(\ref{C_coeff_NNLO}).

\subsubsection{Quark coefficient functions}\label{sec:coeff_quark}
In order to determine the quark coefficient function we perform the calculation of the DY transverse momentum distribution in $k_t$-factorization. While the Born process is $q \bar q \to Z/\gamma^*$, the partonic process that enters $k_t$-factorization is $g q \to Z/\gamma^* q$~\cite{Marzani:2008uh}. As in the Higgs case, we define Mellin moments and we set $N=0$:
\begin{align}\label{DY:qt}
c(M,b)&= \int_0^\infty \d Q_T \, 2 Q_T \,J_0(b Q_T) \,  \frac{h\left(0,M,\xi_p\right)}{\sigma^\text{born}} \nonumber \\&
= \int_0^\infty \d Q_T \, 2 Q_T \,J_0(b Q_T) \, M \int_0^1 \frac{\d z}{z} \int_0^\infty \d \xi \, \xi^{M-1}
\nonumber \\& \times \frac{1}{\sigma^\text{born}}\frac{\d \sigma^\text{off-shell}}{\d \xi_p},
\end{align}
The calculation of the off-shell cross-section is summarized in the Appendix. In the small-$Q_T$ limit, which corresponds to large values of $b$, we find
\begin{align}\label{DY:qt-res}
c(M,b)&=\left(\frac{b_0}{b Q}\right)^{2M} e^{2\gamma_E M} \frac{\Gamma(1+M)}{\Gamma(1-M)} 
\nonumber \\ &\times
\left[ \int_0^\infty \d \zeta \, \zeta^{M-1} f(\zeta) +\ord \left(b^{-2}\right) \right],
\end{align}
where 
\begin{align}\label{DYintegrand}
f(\zeta) &= \frac{\as}{2 \pi} T_R   \nonumber \\ &\times
\begin{cases}
\frac{2}{\zeta}\left( 1-\frac{2-\zeta}{\sqrt{\zeta(4-\zeta)}} \tan^{-1} \frac{\sqrt{\zeta(4-\zeta)}}{2-\zeta}\right)
,\, \zeta<1, \\
\frac{2}{\zeta} \left(1- \frac{\ln \zeta}{2}-\frac{2-\zeta}{\sqrt{\zeta(4-\zeta)}}\tan^{-1} \frac{\sqrt{\zeta(4-\zeta)}}{\zeta}\right)
 , \, \zeta >1.
\end{cases}
\end{align}
Analogous to the gluon-gluon calculation, the $b$-dependent prefactor is absorbed into the PDFs.
We note that the small-$\zeta$ limit of $f$ is actually finite: $f(\zeta)= \frac{\as}{2 \pi} T_R \frac{2}{3} +\ord\left(\zeta \right)$, i.e.\ the first term of $\gamma_{qg}$, as one might expect. 
However, the Mellin integral is singular when $M \to 0$, which corresponds to the collinear region of phase-space. In order to obtain the resummed $\msb$ coefficient function, this singularity has to be subtracted to all orders. The procedure is discussed for deep-inelastic scattering in Ref.~\cite{Catani:1994sq} and for the DY process in Ref.~\cite{Marzani:2008uh}. We have~\footnote{We note that Eq.~(40) of the published version of this paper contained a mistake, namely the collinear subtraction was wrongly multiplied by the Gamma functions prefactor. Once this issue is solved, the resummation is in full agreement with the fixed-order calculation presented in Ref.~\cite{Luo:2019szz}.}
\begin{align}\label{DY:qt2}
&\mathcal{C}_{qg}(M)=
 \tfrac{ e^{2\gamma_E M} \Gamma(1+M)}{\Gamma(1-M)} \int_0^\infty \d \zeta \, \zeta^{M-1} f(\zeta)-\frac{h_{qg}(M)}{M}  \nonumber\\
&=  \tfrac{e^{2\gamma_E M} \Gamma(1+M)}{\Gamma(1-M)} \int_0^\infty \d \zeta \, \zeta^{M} \tilde{f}(\zeta)  \nonumber\\
&\quad \quad \quad\quad \quad \quad  -\frac{h_{qg}(M)-\frac{\as}{2 \pi}T_R\frac{2}{3} \tfrac{e^{2\gamma_E M} \Gamma(1+M)}{\Gamma(1-M)}}{M} ,
\end{align}
with
\begin{align}\label{DYintegrand_tilde}
\tilde{f}(\zeta)&= \frac{\as}{2 \pi} T_R \nonumber \\ &\times
\begin{cases}
\frac{2}{\zeta^2}\left( 1-\frac{2-\zeta}{\sqrt{\zeta(4-\zeta)}} \tan^{-1} \frac{\sqrt{\zeta(4-\zeta)}}{2-\zeta}\right)-\frac{2}{3 \zeta}
, \, \zeta<1, \\
\frac{2}{\zeta^2} \left(1- \frac{\ln \zeta}{2}-\frac{2-\zeta}{\sqrt{\zeta(4-\zeta)}}\tan^{-1} \frac{\sqrt{\zeta(4-\zeta)}}{\zeta}\right)
 , \, \zeta >1,
\end{cases}
\end{align}
and $h_{qg}$ is the impact factor which gives rise to the all-order $\gamma_{qg}$ anomalous dimension, i.e.\ $\gamma_{qg}(N,\as)=R_{\msb}(\gamma_s)\, h_{qg}(\gamma_s)$. The $\msb$ expression for the impact factor $h_{qg}$ and for the all-order (NLL$x$) anomalous dimension $\gamma_{qg}$ are not known in closed-form, but only as a power series to arbitrary accuracy~\cite{Catani:1993rn,Catani:1994sq}, e.g.
\begin{align}\label{gamma_qg}
\gamma_{qg}(N,\as)&=\frac{\as}{2 \pi} T_R \Bigg[\frac{2}{3}+\frac{10}{9} \frac{\as}{\pi} \frac{C_A}{N-1}
\nonumber \\ &+ \frac{28}{27}\left(\frac{\as}{\pi} \frac{C_A}{N-1} \right)^2+\ord \left( \as^3 \right) \Bigg].
\end{align}
The function $\tilde{f}(\zeta)$, Eq.~(\ref{DYintegrand_tilde}) is regular at $\zeta=0$. Therefore, we can expand Eq.~(\ref{DY:qt2}) in powers of $M=\gamma_s$ and integrate term by term. By setting $T_R=\frac{1}{2}$, we obtain
\begin{align} \label{quark_results_qg}
\mathcal{C}_{qg}^{\text{LL}x}(N,\as) &=R_{\msb}(\gamma_s)\,\mathcal{C}_{qg}(\gamma_s)\nonumber \\ &= \frac{\as}{\pi}\frac{1}{12}+ \left(\frac{\as}{\pi}\right)^2 \frac{C_A}{N-1} \left(\frac{43}{108}-\frac{\pi^2}{36} \right) \nonumber \\ &+\ord \left(\as^3 \right), \\
\mathcal{C}_{qq}^{\text{LL}x}(N,\as) &=\frac{C_F}{C_A}\left[R_{\msb}(\gamma_s)\, \mathcal{C}_{qg}(\gamma_s)-\mathcal{C}_{qg}(0)\right]\nonumber \\ &= \left(\frac{\as}{\pi}\right)^2 \frac{C_F}{N-1} \left(\frac{43}{108}-\frac{\pi^2}{36} \right) \nonumber \\ &+\ord \left(\as^3 \right),
\label{quark_results_qq}
\end{align}
The result in Eq.~(\ref{quark_results_qg}) agrees with the known NLO and NNLO expressions Eqs.~(\ref{C_coeff_NLO_DY}) and~(\ref{C_coeff_NNLO_DY}) and the result in Eq.~(\ref{quark_results_qq}) agrees with its NNLO counterpart Eq.~(\ref{C_coeff_NNLO_DY}), in the $N\to 1$ limit. The NLO coefficient $\mathcal{C}^{(1)}_{qq}$ describes the emission of a gluon off a quark line and it is therefore regular in $N=1$ and cannot be predicted from $k_t$-factorization (see discussion in Ref.~\cite{Marzani:2008uh}).

\section{Conclusions and Outlook}\label{sec:conclusions}
In this paper we have discussed the simultaneous resummation logarithms of $Q_T$ and $x$. 
The double-resummed expression that we have obtained, Eq.~(\ref{eq:QT_res_smallx_mellin}), has the same structure as standard $Q_T$ resummation but it contains resummed (and matched) DGLAP anomalous dimensions and coefficient functions. We have computed to LL$x$ all the necessary coefficient functions for Higgs and the DY process, which can be matched to known fixed-order results. 

In order to perform a phenomenological study and assess the impact of small-$x$ resummation on the Higgs and DY $Q_T$ spectrum two further steps are necessary. First, as mentioned in Sec.~\ref{sec:smallx}, the inclusion of subleading running coupling corrections~\cite{Ball:2007ra, Altarelli:2008aj} is required. 
Second, the result Eq.~(\ref{eq:QT_res_smallx_mellin}) has to be matched to the LL$x$ distribution at finite $Q_T$. In particular, for the Higgs case, one needs to compute the $Q_T$ distribution in $k_t$-factorization, in the full theory with finite top mass, in order to obtain the correct small-$x$ limit. The calculation of the DY $Q_T$ spectrum summarized in the Appendix (before Eq.~(\ref{impact_DY_cntd})) is performed at finite $Q_T$ and therefore can be used for matching. However, as mentioned in the Appendix, the analytic evaluation of the integrals at finite $Q_T$ is challenging. 

Both the inclusion of subleading corrections and matching to finite $Q_T$ are work in progress and we look forward to a detailed phenomenological study in the near future.
\begin{acknowledgments}
I~would like to thank Richard Ball, Marco Bonvini, Fabrizio Caola, Stefano Forte and Claudio Muselli for useful discussions and encouragement.
This work is (partly) supported by the U.S.\ National Science Foundation, under grant PHY--0969510, the LHC Theory Initiative.
\end{acknowledgments}

\appendix

\section{Drell-Yan transverse momentum distribution in $k_t$-factorization}\label{app:DY}
In this appendix we describe the calculation of the Drell-Yan transverse momentum distribution in $k_t$-factorization. To this purpose, we make use of the recent results of Ref.~\cite{Forte:2015gve}, which extend the framework of $k_t$-factorization to the case of transverse momentum distributions.

The process of interest here is the gluon-initiated production of a off-shell photon, or a vector boson, with the initial-state gluon taken off its mass-shell. 
For simplicity, we consider only the production of an off-shell photon in a theory with only one quark flavor. The generalization to different quark flavors, as well as the inclusion of the leptonic final states (with either $\gamma^*$ or $Z/W^\pm$) is identical to the invariant-mass distribution case, discussed at length in Ref.~\cite{Marzani:2008uh}.
The LO process, which is relevant for the resummation of the leading non-trivial contribution at small-$x$ is then 
$$g^* (k)\;  q (p) \to  \gamma^*(q) \; q (p'),$$
with 
\begin{align}\label{DYkinematics}
k&= x_1 p_1 + {\bf k}, \quad p=x_2 p_2, \nonumber\\
q&= z_1 x_1 p_1+ (1-z_2) x_2 p_2  + {\bf q}, \nonumber\\ p'&=(1-z_1) x_1 p_1+ z_2 x_2 p_2  + {\bf k}- {\bf q},
\end{align}
with ${\bf a}=(0,\vec{a}_t,0)$, $k^2=-|\kt|^2$, $q^2=Q^2$, $p^2=p'^2=0$. We also find useful to introduce three dimensionless ratios: $\xi=\frac{|\kt|^2}{Q^2}$, $\xi_p=\frac{Q_T^2}{Q^2}$, and $z=\frac{Q^2}{\hat{s}}$. The parton-level transverse momentum distribution can be written as 
\begin{align} \label{off-shell-DY-start}
\frac{\d \sigma^\text{off}}{\d \xi_p} &= \frac{Q^2}{2 \hat{s}}\frac{\hat{s} }{8 \pi^2}\int_0^1  \d z_1 \int_0^1 \d z_2  \int \d^2 \qt |\m|^2
\nonumber\\ &\times
 \delta \left(q^2 -Q^2 \right) \delta  \left(p'^2 \right) \delta \left(|\qt|^2 -Q_T^2\right).
 \end{align}
The matrix element squared was computed in~\cite{Marzani:2008uh}, where the high-energy resummation of the invariant mass distribution was considered. 
The result reads
\begin{align} \label{offshell} |\m|^2 &= -\frac{e_q^2 g_s^2}{N_c}
\Bigg\{\frac{t}{s}+\frac{s}{t}
+Q^2|\kt|^2\left(\frac{1}{s^2} +\frac{1}{t^2} \right)\nonumber \\ &
 + 4\frac{\kt\cdot\qt}{s} -4\frac{Q^2\kt\cdot\qt}{t^2}\left(1-
\frac{\kt\cdot\qt}{|\kt|^2}\right)\nonumber \\ &
+ \frac{2}{st}
\Big[(Q^2+|\kt|^2)(|\kt|^2-2\kt\cdot\qt)
+2(\kt \cdot \qt)^2\nonumber \\ &+|\kt|^2(s-t)\Big] \Bigg \},
 \end{align}
 where the Mandelstam invariants are
 \begin{align}\label{mandelstam}
s &= (p+k)^2 = \hat{s} -|\kt|^2, \nonumber \\
t &= (p-q)^2 = Q^2 - z_1\hat{s}, \nonumber \\
u &= Q^2- |\kt|^2-s-t,
\end{align}
with $\hat{s}= 2 x_1 x_2 p_1 \cdot p_2$, $g_s$ is the strong coupling constant and $e_q$ the quark electric charge. Because we are interested in the $Q_T$ spectrum, we find more useful to organize the calculation of the phase-space integrals following the case of direct photon production~\cite{Diana:2009xv,Diana:2010ef}. Therefore, we use the first $\delta$ function in Eq.~(\ref{off-shell-DY-start}) to perform the integral over the longitudinal momentum fraction $z_2$, obtaining 
\begin{align} \label{off-shell-DY-cntd}
\frac{\d \sigma^\text{off}}{\d \xi_p} &= \frac{1}{16 \pi^2} \frac{z}{2 Q^2}\int_0^1  \frac{\d z_1}{z_1(1-z_1)}  \int_0^{2\pi } \d \phi  \nonumber \\  &\times  \delta  \left(\frac{1}{z}-\frac{1+\xi_p}{z_1}-\frac{\xi+\xi_p-2 \sqrt{\xi \xi_p}\cos \phi}{1-z_1} \right) |\m|^2 \nonumber \\  &\times \Theta \left(\frac{1}{z}-\frac{1+\xi_p}{z_1} \right)  \Theta \left(\frac{1}{z}-\xi \right), 
 \nonumber \\  & \quad \quad \quad \quad \quad \quad  \text{with} \quad z_2=1-z \frac{1+\xi_p}{z_1}.
 \end{align}
 We now compute the impact factor by considering Mellin moments with respect to $\xi$ and $z$, indicating them with $M$ and $N$, respectively. Moreover, because we are interested in the leading singularities we can consider $N=0$. We have
 \begin{align}\label{impact_DY_start}
 h(0,M,\xi_p)&=  M \int_0^\infty \d \xi \, \xi^{M-1} \int_0^1 \d z \int_0^1  \frac{\d z_1}{z_1(1-z_1)}  \nonumber \\  & \times \int_0^{2\pi } \frac{\d \phi}{2 \pi} \, \sigma_0 \,\frac{\as}{2 \pi}\, T_R\, |\widetilde{\m}|^2 
 \nonumber \\  &\times 
 \delta  \left(\frac{1}{z}-\frac{1+\xi_p}{z_1}-\frac{\xi+\xi_p-2 \sqrt{\xi \xi_p}\cos \phi}{1-z_1} \right)  \nonumber \\  & \times \Theta \left(\frac{1}{z}-\frac{1+\xi_p}{z_1} \right)  \Theta \left(\frac{1}{z}-\xi \right), 
 \end{align}
with $\sigma_0=\frac{\pi e_q^2}{N_c Q^2}$ and $|\widetilde{\m}|^2=|\m|^2/\left(\frac{e_q^2 g_s^2}{N_c} \right)$.

We change the integration variable from $z$ to $\rho=1/z$ and we perform this integral using the remaining \mbox{$\delta$-function}, which fixes
\begin{align} \label{rho_value}
\rho= \frac{1+\xi_p-z_1(1-\xi+2 \sqrt{\xi \xi_p}\cos \phi)}{z_1(1-z_1)}.
\end{align}
After some algebra, we can verify that the two $\Theta$ functions in Eq.~(\ref{impact_DY_start}) are always satisfied when $\rho$ is set by Eq.~(\ref{rho_value}).
Performing the remaining integrals analytically becomes now a challenge. However, we note that for the purpose of this paper, we only need the small $Q_T$ behavior of the above distribution, at fixed $k_t$. 
Therefore, we rescale the integration variable $\xi = \xi_p \zeta$ and then we consider only the contribution that does not vanish at small $\xi_p$:
 \begin{align}\label{impact_DY_cntd}
& h(0,M,\xi_p)=  M\,  \xi_p^{M-1} \nonumber \\ &\times \int_0^\infty \d \zeta \, \zeta^{M-1} \, \int_0^1  \d z_1  \int_0^{2\pi } \frac{\d \phi}{2 \pi} \, \sigma_0 \,  \frac{\as}{2 \pi}\, T_R \nonumber \\ &\times \Bigg[
\frac{\zeta  z_1^2-2 \sqrt{\zeta } (2 z_1-1) z_1 \cos \phi +4 (z_1-1) z_1 \cos
   ^2\phi +1}{\left(-2 \sqrt{\zeta } z_1 \cos \phi +\zeta  z_1+1\right)^2}  \nonumber \\  & \quad \quad \quad \quad \quad \quad
   \quad \quad \quad \quad \quad \quad
   \quad \quad \quad  \quad 
   +\ord \left(\xi_p \right)\Bigg].
 \end{align}
The $z_1$ and $\phi$ integrals can now be performed in a closed form, leading to the result quoted in the main text Eqs.~(\ref{DY:qt-res}) and~(\ref{DYintegrand}).

\bibliography{references}

 \end{document}